\documentclass[a4paper,12pt,twocolumn]{article}
\usepackage[cp1251]{inputenc}
\usepackage{epsfig}
\usepackage[english]{babel}
\usepackage{amssymb}

\begin{document}
\onecolumn \vspace{7cm}
\title{Approximate Description of the Mandelbrot Set. \\
Thermodynamic Analogy}
\author{O.B. Isaeva, S.P. Kuznetsov}
%\date{}
\maketitle\begin{center} \emph{Institute of Radio-Engineering and
Electronics of RAS, Saratov Branch, \\ Zelenaya 38, Saratov,
410019, Russia \\ E-mail: IsaevaOB@info.sgu.ru}\end{center}
\begin{abstract}
Analogy between an approximate version of Feigenbaum
renormalization group analysis in complex domain and the phase
transition theory of Yang-Lee (based on consideration of formally
complexified thermodynamic values) is discussed. It is shown that
the Julia sets  of the renormalization transformation correspond
to the approximation of Mandelbrot set of the original map. New
aspects of analogy between the theory of dynamical systems and the
phase transition theory are uncovered.
\end{abstract}
%\tableofcontents

\twocolumn
%***********************************************************************
\section{Introduction} Transition between magnetic and
non-magnetic phase of matter is intensively studied for a long
time~\cite{Rumer,Sinai,Balesku,Peitgen}. At low temperatures,
elementary atomic magnets tend to be parallel providing coherent
(magnetic) phase of the system. With increase of the temperature,
thermal fluctuations break this order drastically. If the
temperature is higher than the phase transition critical point
(Curie point), then chaotic orientation of elementary magnets
takes place (non-magnetic phase). However, near the critical
point, the elementary magnets still keep order on certain range of
distances and over certain time intervals, which decrease with
growth of the temperature. Thus, at the critical point, areas of
coherence and areas of thermal fluctuations coexist. Areas of
fluctuations in the magnetic phase or areas of coherence in the
non-magnetic phase may be distingueshed only within a certain
range of scales. For large enough scales the system looks
completely ordered (as at zero temperature) or chaotic (as at
infinite temperature). If one considers the same system at a
definite temperature but within various scales, it looks as the
system at different temperatures. The transformation of scales
corresponds to the temperature renormalization.

Let we have a cubic lattice of $N$ atoms with interatomic distance
$a$ at temperature $T$. If one considers this system at rough
enough scale, at which the elementary cell has the characteristic
size $na$ and contains $n^3$ atoms, the system looks as a lattice
of $N/n^3$ atoms at the renormalized temperature $T'=R(T)$ ($R$ is
a function of renormalization). At the moment of the phase
transition, one can observe the thermal fluctuations within any
scales. These fluctuations are self-similar. The idea of
self-similarity has been offered by Kadanoff~\cite{Kadanoff} and
further advanced by Wilson to the renormalization
method~\cite{Wilson} (see also~\cite{Sinai}).

Concepts of universality and scaling near the critical point and a
renormalization group (RG) method have been transferred to
nonlinear dynamics by Feigenbaum~\cite{Feigenbaum1,Feigenbaum2}.
The analogy between transition to chaos and phase transitions
consist in increase of temporal scales with approach to a critical
point, self-similarity, an opportunity of application of RG
analysis, and in existence of universal critical indexes and
scaling factors, which are determined by the most general
requirements to the type of
system~\cite{Shuster,Haken,Kuznetsov1,Kuznetsov2}.

In 1952 Lee and Yang advanced an approach in the phase transition
theory based on consideration of analytical properties of some
thermodynamic values, such as partition function and free energy,
depending on the temperature considered formally as a complex
variable~\cite{Yang1,Yang2}.

The partition function is defined as follows:
\begin{equation}\label{1}\begin{array}{c}
Z_N (T,H) = \sum\limits_{|s_i|} \exp \left( - \frac{E(s_i)}{k
T}\right) = \\ =\sum\limits_{|s_i|} \exp \left({\frac{{J}}{{k
T}}}{\sum\limits_{i,j} {s_{i}} } s_{j} + {\frac{{H}}{{k
T}}}{\sum\limits_{i} {s_i} } \right),
\end{array}
\end{equation}
where $E$ is a total energy of a configuration of the system, $H$
is an external magnetic field, $T$ is a temperature, $k$ is the
Boltzmann constant, $s_i$ - is the spin variable, taking place at
$i$-th cell of the lattice. Spins interact via magnetic field with
the nearest neighbours ($J>0$ corresponds to ferromagnetic
interaction, $J<0$ -- antiferromagnetic). Partition function can
be transformed to
\begin{equation}\label{2}
Z_N= e^{-N/2}\sum\limits_{n=0}^{N} p_nz^n,
\end{equation}
by the variable change $z=\exp(-2H/kT)$. It is known, that at the
phase transition the divergence of the thermodynamic potential
(free energy)
\begin{equation}\label{3}
f(H,T)=\lim \limits_{N\rightarrow\infty} -\frac{kT}{N}\ln Z_N
\end{equation}
takes place.

Obviously, zeros of the partition function (roots of the algebraic
equation $Z_N=0$) are the candidates for the phase transition
points. However, these roots should be positive and real, which is
impossible as all factors $p_n>0$. In Yang-Lee theory, the
temperature and the magnetic field are considered as complex
variables~\cite{Peitgen,Yang1,Yang2}. Complex zeros of the
partition function (called the Yang-Lee zeros), gather near the
real axis. It becomes possible only in the thermodynamic limit,
the asymptotics of infinite number of atoms. When the number of
atoms is finite, there is a finite number of zeros in the complex
plane. With increase of number of atoms, the set of zeros becomes
dense and nestles to the real axis more and more. In the
thermodynamic limit the Yang-Lee zeroes form a fractal set, that
cross the real temperature axis at the point of the phase
transition. To find the Yang-Lee zeros it is appropriate to apply
the Wilson renormalization method, which corresponds to a
consecutive decrease of the number of degrees of freedom of the
partition function. It is necessary to find the transformation of
the $N'$-particle partition function to the $N$-particle one
($N'<N$). This transformation is not reversible. Thus, to get
zeros of $Z_N$ it is necessary to have zeros of $Z_{N'}$, and then
to construct their backward images using the RG transformation.
Having repeated this procedure many times one comes finally to the
trivial two-atom partition function. If zeros of $Z_2$ belong to a
basin of attraction of a fixed point of the RG transformation,
then zeros of $Z_N$ in the thermodynamic limit will coincide to
the boundary of the basin of attraction in the complex plane. As
shown by Derrida~\cite{Derrida}, this boundary, which is the Julia
set of the RG transformation, is identical to the set of Yang-Lee
zeros. Unfortunately, the renormalization transformation may be
performed analytically  only for simple class of models -- the
hierarchical lattices (for example, one and two-dimensional model
with Ising spins~\cite{Onsager}, hierarchical lattices with Potts
spins~\cite{Ananikian1,Ananikian2}).

Yang-Lee theory appears to be fruitful for understanding the phase
transitions. It seems that a similar approach to the analysis of
transition to chaos in dynamical systems would be useful for
deeper understanding of the analogy with the phase transitions,
and for development of the new criteria describing complexity of
behavior of nonlinear systems. In the present paper we develop
such approach basing on the approximate RG analysis of transition
to chaos through the period multiplication bifurcations.

\section{Approximate RG analysis}
In 1978 Feigenbaum discovered universality of cascade of the
period-doubling bifurcations and described it on a basis of the RG
method. The simplest example representing the Feigenbaum
universality class is quadratic map
\begin{equation}\label{4}
x_{n+1}=f_{\lambda}(x_n)= \lambda-x_n^2,
\end{equation}
where $x$ is a real dynamical variable, and $\lambda$ is a real
parameter. This map has fixed points, which can be found as roots
of the equation $x_*=\lambda-x_*^2$. If $\lambda>\Lambda_1=3/4$, a
cycle of period $1$ (that is the fixed point) loses its stability.
This $\Lambda_1$ is the parameter, at which the first
period-doubling bifurcation occurs (the multiplier of the fixed
point is $\mu=f'_{\lambda}(x_*)=-2 x_*=-1$). Values of $\lambda$
for sequent bifurcations can be found by means of approximate RG
method~\cite{Landau}. Let's apply the map (4) two times:
\begin{equation}\label{5}
x_{n+2}=\lambda-\lambda^2+2\lambda x_n^2-x_n^4
\end{equation}
and neglect the last term, the fourth power of $x_n$. Then, by the
scale transformation
\begin{equation}\label{6}
x_n \rightarrow x_n/ \alpha_0,\qquad \alpha_0=-2\lambda.
\end{equation}
this map can be rewritten in the form $x_{n+2}=\lambda_1-x_n^2$,
which differs from (4) only by renormalization of $\lambda$
\begin{equation}\label{7}
\lambda_1=\varphi(\lambda)=-2\lambda(\lambda-\lambda^2).
\end{equation}
Thus, the operator of evolution for the double interval of
discrete time can be reduced to the original operator by the
renormalization transformation~(7). Repeating this procedure with
scale factors $\alpha_1=-2\lambda_1,...$, one can obtain a
sequence of the same form
\begin{equation}\label{8}
x_{n+2^m}=\lambda_m-x_n^2, \qquad
\lambda_m=\varphi(\lambda_{m-1}).
\end{equation}

Fixed points of these maps correspond to the $2^m$-cycles of the
original map ($m=1,2,3,...$). It is easy to see, that all these
cycles, as well as the fixed point of the map~(4), become unstable
at $\lambda_m=\Lambda_1=3/4$. Solving a chain of the equations
\begin{equation}\label{9}
\Lambda_1=\varphi(\Lambda_2), \Lambda_2=\varphi(\Lambda_3),
...,\Lambda_{m-1}=\varphi(\Lambda_m)
\end{equation}
we get the corresponding sequence of bifurcation values of
parameter $\lambda$ (with $\lambda\approx\Lambda_m$ the
$2^m$-cycle of~(4) arises). From iteration diagram of Fig.1 it is
evident, that this sequence converges with $m\rightarrow \infty$
to a definite limit $\Lambda_{\infty}$, the fixed point of the RG
transformation. It satisfies the equation
$\Lambda_{\infty}=\varphi(\Lambda_{\infty})$, thus
$\Lambda_{\infty}=(1+\sqrt{3})/2\approx 1.37$. The scaling factors
also converge to the limit: $\alpha_m\rightarrow \alpha$, where
$\alpha=-2\Lambda_{\infty}\approx 2.74$. The multipliers (Floquet
eigenvalues of the $2^m$-cycles) converge to the universal value
$\mu_m\rightarrow \mu=\sqrt{1-4\Lambda_{\infty}}\approx -1.54$.

From transformation (8) it is possible to obtain the law of
convergence of the bifurcation sequence:
\begin{equation}\label{10}\begin{array}{l}
\Lambda_m=\varphi(\Lambda_{\infty})+\varphi'(\Lambda_{\infty})(\Lambda_{m+1}-\Lambda_{\infty})= \\
\qquad\qquad\qquad\qquad
=\Lambda_{\infty}+\delta(\Lambda_{m+1}-\Lambda_{\infty})
\end{array}\end{equation}
where $\delta=\varphi'(\Lambda_{\infty})=4+\sqrt{3}\approx 5.73$
is a constant, characterizing the convergence to the critical
point.

In table~1 we summarize the values of critical indexes (critical
point, scale factor, parameter scaling constant and universal
multiplier) obtained by means of the exact and approximate RG
analysis. The correspondence between them is well enough.

\begin{table}
  \centering
\caption{Rigorous and approximate values of Feigenbaum critical
indexes}
\begin{tabular}{|c|c|c|}
\hline
    & Rigorous & Approximate \\
\hline
  $\lambda_*$ & 1.401 & 1.37 \\
  $\alpha$ & -2.802 & -2.74 \\
  $\delta$ & 4.669 & 5.73 \\
  $\mu$ & -1.569 & -1.54 \\ \hline
\end{tabular}

\end{table}

\begin{figure}[h!]% Fig.1.
\centering
\includegraphics[width=0.33\textwidth,keepaspectratio]{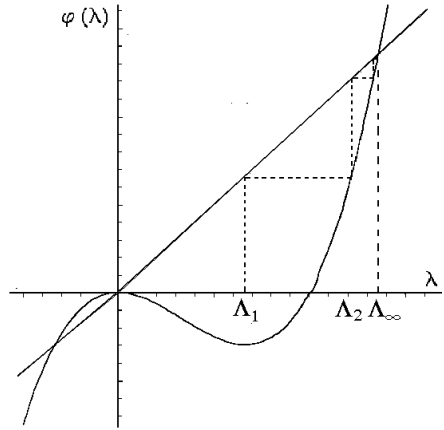}
\caption{Iteration diagram of the RG transformation (7). Dashed
line designates the backward iterations starting at the first
period doubling bifurcation point ($\Lambda_1=3/4$) and mapping to
the further bifurcation points $\Lambda_m$.}
\end{figure}

\section{Complex RG transformation and its Julia set: The critical phase transition line}
Let us regard the variable $x$ and parameter $\lambda$ of map (4)
as complex numbers. Then it becomes clear, that Feigenbaum's
universal scaling laws correspond to a special case of general
scaling properties of Mandelbrot sets. The Mandelbrot set shown in
Fig.2, can be determined as~\cite{Peitgen}
\begin{equation}\label{11}
M=\{\lambda \in {\bf C}:
\lim\limits_{n\rightarrow\infty}f_{\lambda}^n(0) \neq\infty \}.
\end{equation}
Thus, $M$ is the set of values of complex parameter, for which the
trajectories launched from the extremum of the map (4)
$0\rightarrow\lambda\rightarrow\lambda-\lambda^2 \rightarrow,,,$
remain in a finite domain. Mandelbrot set includes parameter
values corresponding to periodic trajectories of the map (the main
cardioid and a number of the round "leaves", painted by gray
color), and a set of parameter values corresponding to bounded
chaotic dynamics (black fractal pattern). Feigenbaum's cascade of
period-doublings takes place along the real axis (see the picture
of the bifurcation tree in the Fig.2).

However, in the complex plane one can observe many other
accumulation points of other bifurcation cascades. One of such
points, with scaling properies distinct from the Feigenbaum laws
and intrinsic exclusively for complex analytical
maps~\cite{Cvitanovic,Isaeva} is associated with the cascade of
period tripling~\cite{Golberg}. Also, there exist critical points
connected with cascades of period quadrupling, period 5-tupling,
etc.

\begin{figure}[h!]% Fig.2.
\includegraphics[width=0.36\textwidth,keepaspectratio]{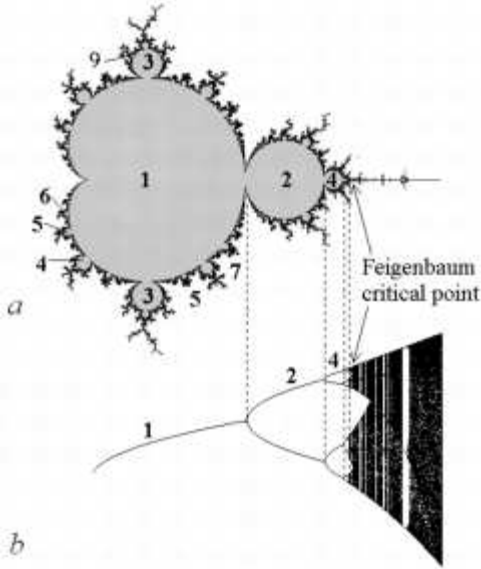}
\caption{Correspondence between the Mandelbrot set (a) and the
Feigenbaum bifurcation tree (b). Mandelbrot set is built up on a
plane $({\rm Re}\lambda, {\rm Im}\lambda)$ for the complex
quadratic map (4). Bifurcation tree is shown in the plane
$(\lambda, x_n)$ for the map (4). Gray color in panel (a)
indicates areas of periodic dynamics (the periods of cycles are
marked); black color designates points corresponding bounded in
the phase space chaotic dynamics; white color means escape to
infinity.}
\end{figure}

Thus, in the complex domain, a number of new scenarios of
transition to chaos and other critical phenomena occurs. Let us
consider the approximate RG transformation generalized on a
complex plane
\begin{equation}\label{12}
\lambda_{n+1}=\varphi(\lambda_n)=-2\lambda_n(\lambda_n-\lambda_n^2).
\end{equation}

Starting from any point of a complex plane the trajectory of the
map approaches the attractor at $\lambda=0$ or escapes to the
attractor at infinity. The borderline of these basins of
attraction is the Julia set $J$ of the of the complex
mapping~(12)~\cite{Peitgen}. Let us call it the {\it critical set}
by analogy with the theory of phase transitions, where it is
interpreted as the critical phase border. Julia set $J$ is
determined as the border of the set
\begin{equation}\label{13}
P=\{\lambda \in {\bf C}:
\lim\limits_{n\rightarrow\infty}\varphi^n(\lambda) \neq\infty \}.
\end{equation}

In Fig.3 a diagram ot the complex plane of initial values of
$\lambda$ of RG transformation (12) is represented. Black color
indicates the basin of attraction $P$, that is the area bounded by
the Julia set $J$. Shades of gray color designate areas of
different time of escape (i.e. dynamic distance up to attractor at
infinity),  i.e. the sets of points with various number of
iterations necessary to escape out of a large enough ball.

It is evident that the Julia set in Fig.3 is quite similar to the
Mandelbrot set of the map (4) (Fig. 2). Similarity between them
can be explained as follows. Indeed, the escape of $\lambda$ to
infinity with iterations of RG transformation means the escape to
infinity of iterations of the map (4). Distinctions can be
explaned by approximate character of RG transformation. Also, it
is necessary to take into account that transformation (12)
describes properties only for the period-doubling cascade of
period-doublings, thus providing the similarity of sets $M$ and
$J$ only near the Feigenbaum point. (The asymptotic similarity of
the area of an analyticity of exact solution of the RG equations
and Mandelbrot set near the Feigenbaum point was discussed earlier
in works~\cite{Nauenberg,Buff,Wells}.) It is interesting to
generalize these results for other sequences of
period-multiplication bifurcation cascades.

\begin{figure}[h!]% Fig.3.
\centering
\includegraphics[width=0.3\textwidth,keepaspectratio]{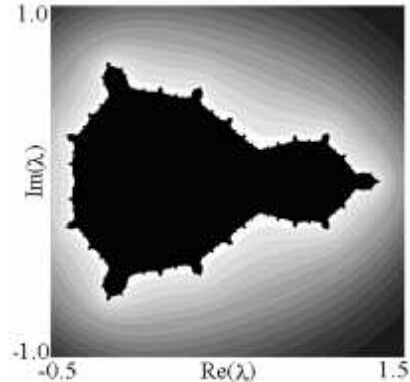}
\caption{Complex plane of initial values of $\lambda$ for RG
transformation (12) is represented. Black color designates the
basin of attraction of origin bounded by the Julia set. Shades of
gray color mark the area escape to infinity. Light colors
correspond to the larger dynamical distance from the attractor at
infinity.}
\end{figure}

\section{Complex cycles of RG transformation}
Let us describe procedure of visualisation of critical set $J$
with the help of a backward iterations. This methode is based on
known fundamental properties of the Julia sets of analytical maps
\cite{Peitgen}:

If there is a repeller point $\lambda_*$ for the map
$\lambda_{n+1}=\varphi(\lambda_n)$, then the set
\begin{equation}\label{14}
J'=\{ \lambda: \varphi^n(\lambda)= \lambda_*, n=1,2,...\},
\end{equation}
is dense in $J$. The fixed point can be found directly from an
iterative polynomial, as a root of the equation
$\varphi(\lambda_*)=\lambda_*$. If the fixed point satisfies a
condition $|\varphi'(\lambda_*)|>1$, then it is repeller.

Julia set is invariant both in respect to forward and backward
iterations. Backward orbit of any point laying in some basin of
attraction approaches closer to a border of this basin, i.e. to
the Julia set.

Repeller points in our case are known to be
$\lambda_*=(1\pm\sqrt{3})/2$. The backward transformation also can
be found analytically by solving equation (12) in respect to the
variable $\lambda_n$ (we easily do it with the help of software
packet {\it Mathematica})
\begin{equation}\label{15}\begin{array}{c}
\lambda_n^1=\frac{1}{3}+\frac{2^{2/3}}{3(4+27\lambda_{n+1}+3\sqrt{3\lambda_{n+1}(8+27\lambda_{n+1})})^{1/3}}+
\\
\frac{(4+27\lambda_{n+1}+3\sqrt{3\lambda_{n+1}(8+27\lambda_{n+1})})^{1/3}}{3\times
2^{2/3}}, \\ \lambda_n^{2,3}=\frac{1}{3}-\frac{1\pm\sqrt{3}
i}{3\times
2^{2/3}(4+27\lambda_{n+1}+3\sqrt{3\lambda_{n+1}(8+27\lambda_{n+1})})^{1/3}}
- \\ \frac{(1\pm\sqrt{3} i)(
4+27\lambda_{n+1}+3\sqrt{3\lambda_{n+1}(8+27\lambda_{n+1})})^{1/3}}{6\times
2^{2/3}}.
\end{array}
\end{equation}

Also the Julia set $J$ can be regarded as a set of all unstable
cycles of every possible periods $n$~\cite{Peitgen,Widom,Jensen}.
With increase of $n$ this set becomes more and more dense, and at
$n\rightarrow\infty$ the distribution of the points on the complex
plane yields the Julia set. Numerical calculation of an unstable
cycle of period $n$ is connected with constructing of every
possible periodic sequences $\varepsilon_1, \varepsilon_2,...,
\varepsilon_n, \varepsilon_1, \varepsilon_2,...,
\varepsilon_n,...$, where $\varepsilon_i$ takes values $1$, $2$ or
$3$, which correspond to a choice of a root $\lambda^1$,
$\lambda^2$ or $\lambda^3$ at the $i$-th backward iteration of
(15). As a result of iteration procedure, the sequence of values
of $\lambda$ converges to a cycle of period not exceeding $n$.

So, one of the unstable fixed points of RG transformation (12) is
situated on the real axis and corresponds to the Feigenbaum
transition to chaos. Beside it, there are unstable cycles of any
other period. Elements of a cycle of period $n$ of RG
transformation (12) correspond to the accumulation points of
bifurcations of period $2^n$-tupling on the Mandelbrot set. This
fact as well explains the similarity of the critical Julia set and
the Mandelbrot set for the original map.

The sets of unstable cycles of periods $6-11$ are shown in Fig.4
(a-f). One can see that with increase of $n$, the full picture of
critical set becomes more and more distinguishable. The analogy
between elements of the cycles and Yang-Lee zeros, and also
between limit of the infinite period of cycles $n$ and a
thermodynamic limit in the theory of phase transitions is evident.

\begin{figure}[h!]% Fig.4.
\centering
\includegraphics[width=0.4\textwidth,keepaspectratio]{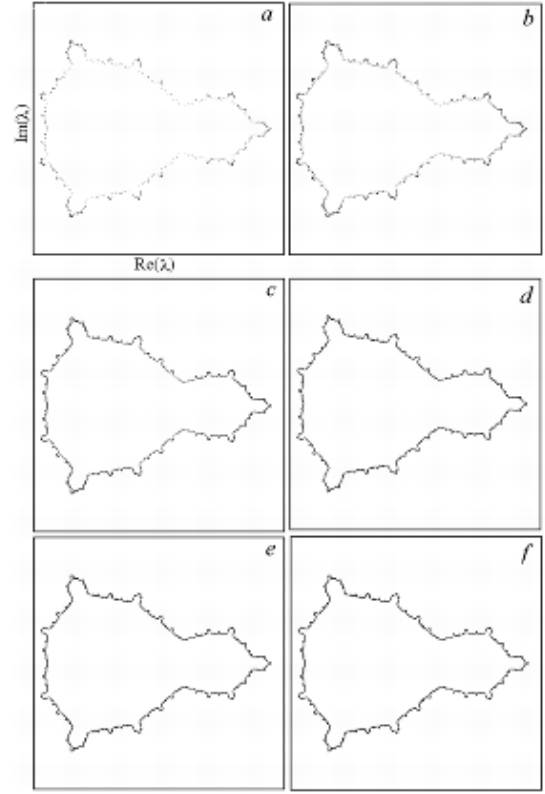}
\caption{Sets of all unstable cycles with periods $n=6$ (a), $7$
(b), $8$ (c), $9$ (d), $10$ (e), $11$ (f) obtained by backward
iterations of RG transformation (12).}
\end{figure}

\section{Electrostatical analogy -- the potential of critical set and its properties near the transition to chaos}
The phase transition critical point corresponds to intersection of
the real axis with the critical line in the complex plane, formed
in the thermodynamical limit by the set of zeros of partition
function. This critical point should provide the jump of a
derivative of the thermodynamic free energy (in the case of I-type
phase transition) or of the second derivative (in the case of the
II-type phase transition). Let us notice that the distribution of
free energy depending on temperature can be interpreted as the
distribution of the electrostatic potential created in 2D space by
the set of charges, located at the partition function zeros, and
in thermodynamic limit -- the potential of the corresponding
limiting distribution of charge on the critical Julia set.

The electrostatic potential for critical Julia set can be
calculated from the Hubbard and Douady~\cite{Peitgen} formula
\begin{equation}\label{16}
U=\frac{1}{3^n}\lim\limits_{n\rightarrow\infty}
\ln|\varphi^n(\lambda)|.
\end{equation}
This formula arises from reasons of existence of conformal mapping
of the Julia set to a circle. The factor is not so important. In
our case it is defined by the third order of the map (12): By one
iteration, the potential of the point increase three times. At
infinity, the potential of the Julia set coincides with potential
of the charged disk $U_0=(1/3^n)\ln|z|$.

We have examined distribution of the electrostatic potential and
its derivatives, both in complex area, and on the real axis. In
Fig.5 the set of several equpotential lines of homogeneously
charged 2D object limited by Julia set of map (12) are shown. This
figure characterizes the distribution of potential in the complex
plane. In Fig.6 the distribution of the electrostatic normal
component of a field (derivative of the potential) along the real
axis is shown. The fact of existence of the jump of the derivative
of the potential at the critical point proves to be true. Near the
other point of crossing of the critical set with the real axis the
first derivative of potential behaves continuously, but has a
break providing jump of the second derivative. This point has to
be interpreted as the II-type phase transition. As appears, the
dependence of potential close to the Feigenbaum critical point
behave asymptotically as $U({\rm Re}\lambda)= ({\rm
Re}\lambda-{\rm Re}\lambda_*)^\gamma$, where the power factor
$\gamma\simeq 0.63$ (see Fig. 7).

\begin{figure}[h!]% Fig.5.
\centering
\includegraphics[width=0.4\textwidth,keepaspectratio]{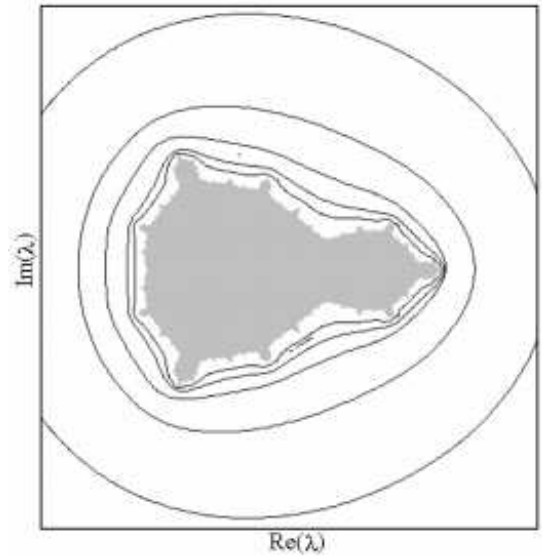}
\caption{Julia set of the map (12) and its equpotential lines,
demonstrating distribution of electrostatic field at the areas of
escape to infinity on the complex plane.}
\end{figure}

\begin{figure}[h!]% Fig.6.
\centering
\includegraphics[width=0.43\textwidth,keepaspectratio]{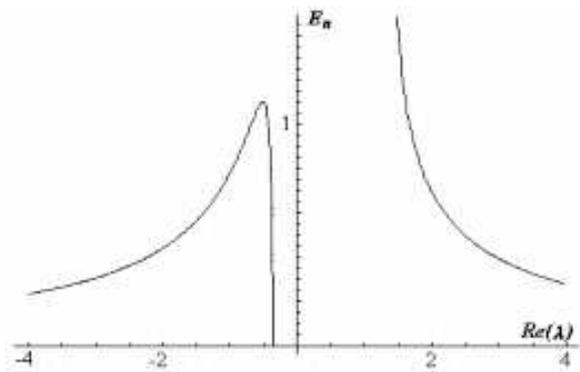}
\caption{Distribution of normal component of intensity of the
electrostatic field of the critical set along the real axis
$\lambda$.}
\end{figure}

\begin{figure}[h!]% Fig.7.
\centering
\includegraphics[width=0.43\textwidth,keepaspectratio]{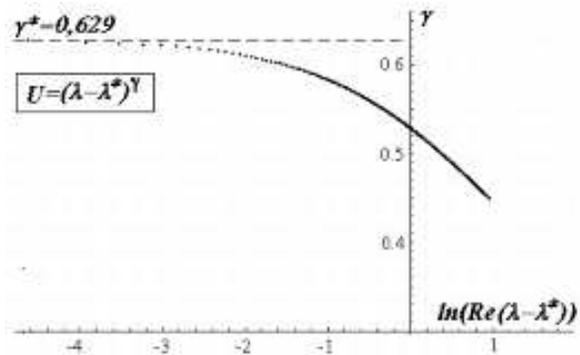}
\caption{Distribution of inclination of a double logarithmic curve
$\ln U({\rm Re} \lambda)$ vs $\ln ({\rm Re} \lambda-{\rm Re}
\lambda_*)$ along real axis $\lambda$ near the Feigenbaum critical
point $\lambda_*$.}
\end{figure}

According to these results, we can assert that the potential of
the Julia set of the RG transformation is a certain criterion of
ordering of chaotic dynamics near to the point of transition to
chaos, like the free energy is connected with the order parameter
(magnetization) of thermodynamic system. One of known analogue of
the order parameter is Lyapunov exponents~\cite{Sinai}. However
sometimes it is more effective to use the electrostatic potential
for the description of dynamics, for example when the trajectories
escape to infinity.

%===========================================================%
\section{RG analysis of period-multiplication bifurcations cascades}
Let us generalize the approximate renormalization procedure for
other bifurcation cascades, distinct of that of Feigenbaum.
Namely, we consider the period-multiplications such as
period-tripling, period-quadrupling etc. intrinsic to complex
analytic maps.

To derive the RG transformation of the period-tripling
(quadrupling) critical point, one must apply the original
map~(\ref{4}) three (four) times and represent the result as a
Taylor series up to quadratic terms in $x_n$:
\begin{equation}\label{17}
\begin{array}{l}
x_{n+3}=f(f(f(x_n)))=\lambda-(\lambda-\lambda^2)^2-\\
\qquad 4\lambda(\lambda-\lambda^2)x_n^2+O(x_n^4 ),
\end{array}
\end{equation}
\begin{equation}\label{18}
\begin{array}{l}
x_{n+4}=f(f(f(f(x_n))))=\lambda-(\lambda-(\lambda-\lambda^2)^2)^2+ \\
\quad
8\lambda(\lambda-\lambda^2)(\lambda-(\lambda-\lambda^2)^2)x_n^2+O(x_n^4).
\end{array}
\end{equation}
From~(\ref{17}) and~(\ref{18}) one can obtain the parameter
renormalization transformations for period
tripling~(Eq.~(\ref{20})) and period-quadrupling~(Eq.~(\ref{22}))
respectively:
\begin{equation}\label{20}
\lambda'\rightarrow
4\lambda(\lambda-\lambda^2)(\lambda-(\lambda-\lambda^2)^2),
\end{equation}
\begin{equation}\label{22}
\lambda'\rightarrow
-8\lambda(\lambda-\lambda^2)(\lambda-(\lambda-\lambda^2)^2)(\lambda-(\lambda-(\lambda-\lambda^2)^2)^2)
\end{equation}
and respective renormalization factors are expressed as:
\begin{equation}\label{19}
\alpha=4\lambda(\lambda-\lambda^2),
\end{equation}
\begin{equation}\label{21}
\alpha=
-8\lambda(\lambda-\lambda^2)(\lambda-(\lambda-\lambda^2)^2).
\end{equation}

\begin{figure}%
\centerline{
\includegraphics[width=0.23\textwidth,keepaspectratio]{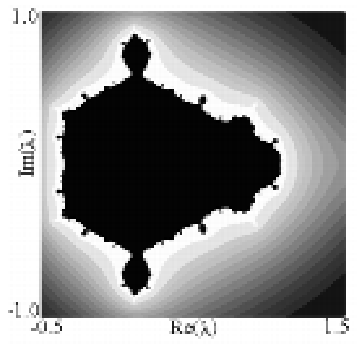}
\includegraphics[width=0.237\textwidth,keepaspectratio]{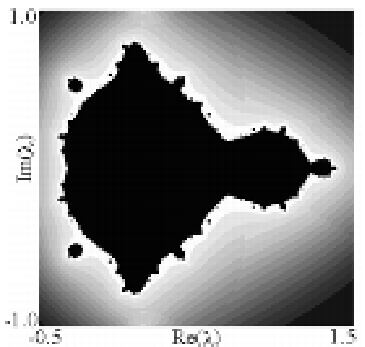}}
\centerline{$a$\hspace{4cm}$b$} \centerline{
\includegraphics[width=0.23\textwidth,keepaspectratio]{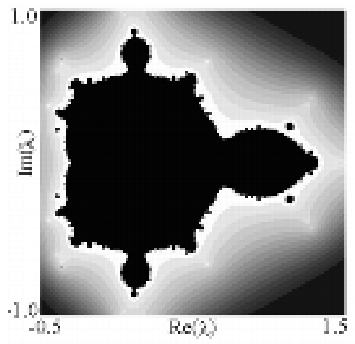}
\includegraphics[width=0.237\textwidth,keepaspectratio]{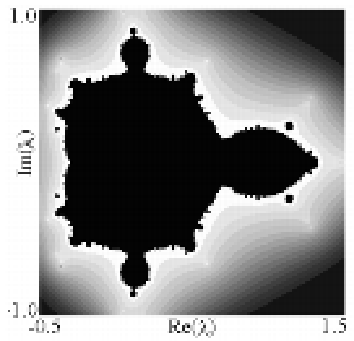}}
\centerline{$c$\hspace{4cm}$d$}

\caption{Charts of initial values of parameter $\lambda$ of RG
transformations for different order of bifurcation cascade
$N=3$~($a$), $N=4$~($b$), $N=2\ast 3$~($c$), $N=2\ast 2 \ast
3$~($d$)}
\end{figure}
\begin{figure}%
\centerline{
\includegraphics[width=0.47\textwidth,keepaspectratio]{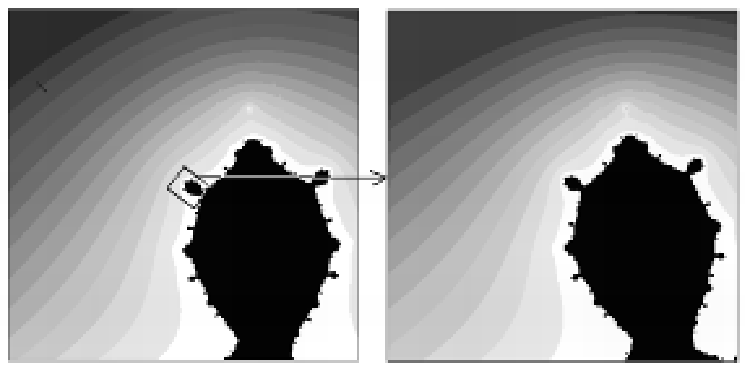}}
\centerline{$a$} \centerline{
\includegraphics[width=0.47\textwidth,keepaspectratio]{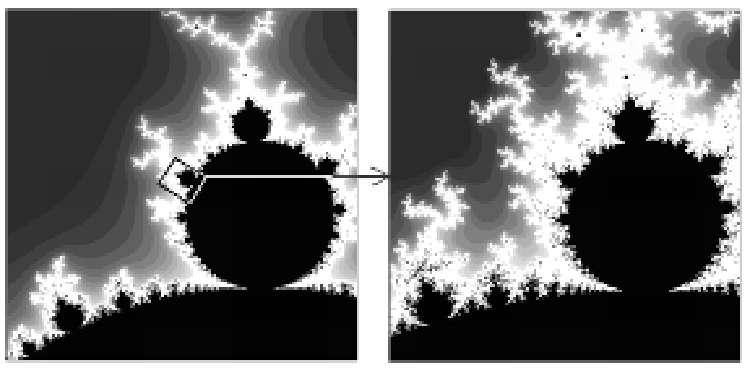}}
\centerline{$b$}

\caption{Scaling of the Julia set of approximate RG
transformation~($a$) and Mandelbrot set of the map~(\ref{4})~($b$)
near the critical point of period-tripling bifurcations cascade.}
\end{figure}

Julia sets of the transformations~(\ref{20}) and~(\ref{22}) are
shown in figure~8~($a$,$b$). Critical values of parameter
$\lambda$, scaling factors, parameter scaling constants and
critical multipliers are numerically calculated. It is evident
that the critical indices are represented by complex numbers. It
is possible to compare them with the data of the numerical RG
analysis~\cite{Cvitanovic} (see Tables~2 and~3). For example, let
us consider one of the fixed points of the
transformation~(\ref{20}), which corresponds to the
Golberg-Sinai-Khanin point of the period-tripling
accumulation~\cite{Isaeva,Golberg}. Figure~9~($a$) demonstrates
scaling properties of the critical set of the
transformation~(\ref{20}) near the approximate period-tripling
accumulation point. On panel~($b$) an illustration of the scaling
properties of the Mandelbrot set near the Golberg-Sinai-Khanin
point is shown. A small fragment of the critical set~(fig.9$a$)
(Mandelbrot set~(fig.9$b$)) by the multiplication to the
approximate (rigorous) scaling constant $\delta$ transforms to a
picture in the right column. As the constant $\delta$ is a complex
number, this transformation includes a scale change and a
rotation.

\begin{table}\centering
\caption{Rigorous and approximate values of critical indexes for
the period-tripling accumulation point}
\begin{tabular}{|c|c|c|}
\hline
 &Rigorous&Approximate\\
\hline
$\lambda^*$ & 0.024+0.784i  & 0.025+0.792i\\
$\alpha$   & -2.097+2.358i & -2.317+2.147i\\
$\delta$   & 4.600+8.981i  & 7.078-7.624i\\
$\mu_c$      & -0.476-1.055i & -0.493-1.062i\\

\hline
\end{tabular}
\end{table}

\begin{table}\centering
\caption{Rigorous and approximate values of critical indexes for
the period-quadrupling accumulation point}

\begin{tabular}{|c|c|c|}
\hline
 &Rigorous&Approximate\\
\hline
$\lambda^*$ & -0.310+0.495i & -0.314+0.493i\\
$\alpha$   & -1.131+3.260i & -1.238+2.913i\\
$\delta$   & -0.853-18.110i  & 3.041-13.984i\\
$\mu_c$      & 0.063-1.053i & 0.070-1.060i\\

\hline
\end{tabular}
\end{table}

Let us generalize the approximate RG analysis to the period
$N$-tupling bifurcations cascades for an arbitrary $N$. By
induction, the expression for the scaling factor and the parameter
renormalization transformation look as follows:
\begin{equation}\label{23}
\alpha=(-2)^{N-1}\prod\limits_{i=1}^{N-1}f^i_\lambda(0),
\end{equation}
\begin{equation}\label{24}
\lambda'\rightarrow(-2)^{N-1}\prod\limits_{i=1}^{N}f^i_\lambda(0).
\end{equation}
Obviously, the critical Julia sets of the RG
transformations~(\ref{24}) with increasing $N$ approximate the
fractal properties of the Mandelbrot set more and more precisely.
For example, one of the fixed points of the RG transformation of
period-quadrupling takes place on the real axes at
$\lambda^*=1.396$. This point corresponds to the Feigenbaum
critical point $\lambda^*=1.401$ and is approximated in this
renormalization scheme more precisely than previous estimate
$\lambda^*=1.37$.

It is worth noting that for the better approximation of the
Mandelbrot set by Julia sets of the RG transformation, for a large
period multiplication the factor $N$ (increasing to infinity) must
be a composite number, that is $N=2\cdot3\cdot...$ Then the
considered RG transformation will describe a number of different
bifurcations cascades (see figure~8).

Let us construct a map, which allows an RG transformation of order
$N+1$ using RG transformations of low orders. It yields
\begin{equation}\label{25}
\begin{array}{l}
\lambda_{N+1}=(-2)\lambda_{N}\ast \\

\ast(\frac{\lambda_{N}}{-2\lambda_{N-1}}-(\frac{\lambda_{N}}{-2\lambda_{N-1}})^2-(\frac{\lambda_{N-1}}{-2\lambda_{N-2}})^2),\\
\quad \lambda_1=f_\lambda(0)=\lambda, \lambda_0=1,
\lambda_{-1}=\infty.
\end{array}
\end{equation}
Dynamics of such a map with $N\rightarrow\infty$ includes
description of all period $N$-tuplings cascades. Figure~10 shows
the plane of initial values of $\lambda_1=\lambda$. One can see
that it precisely corresponds to the Mandelbrot set.

\begin{figure}%
\centerline{
\includegraphics[width=0.3\textwidth,keepaspectratio]{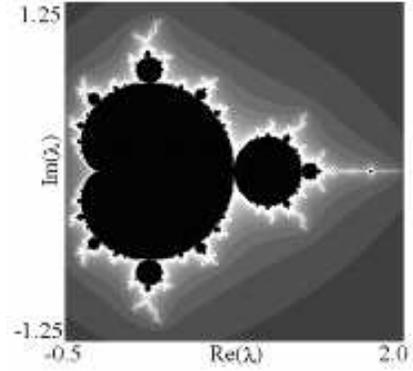}}
\caption{Julia set of the approximate RG transformation of
bifurcation cascade of infinite order $N\rightarrow\infty$.}
\end{figure}

\section{Conclusion}
In the present paper the complex variable version of approximate
RG method for period-doubling bifurcation cascade is considered.
It is shown, that the Julia set of renormalization transformation
of parameter of complex logistic map is the approximate version of
Mandelbrot set of this map. This similarity is explained by the
fact, that Julia set is a set of every possible unstable cycles of
RG transformation, and elements of these cycles correspond to
accumulation points of various bifurcations of period $2^n$
multiplication ($n\rightarrow\infty$), located on the boundary of
Mandelbrot set dense everywhere.

Approximate RG analysis is generalized to the case of different
cascades of period-multiplication bifurcations (for example
period-tripling, period-quadrupling etc.), which is peculiar for
the complex analytic maps. It is shown that with increase of order
of the bifurcation cascade the similarity between Mandelbrot set
and Julia set of renormalization transformation becomes more
clear.

The obtained outcomes are interpreted in a view of analogy with
theory of phase transitions, namely the Yang-Lee theory, based on
investigation of properties of thermodynamic values depending on
complex temperature. It is necessary to note, that the elements of
considered unstable cycles of RG transformation which  is
equivalent to points of Julia set of this transformation, in
thermodynamic analogy corresponds to so-called Yang-Lee zeros,
defining in a thermodynamic limit the borderline of phase
transition. It is shown, that at the points of transition to chaos
the jump of an electrostatic field of a critical Julia set is
observed. Within the framework of considered analogy, the jump of
a derivative of a free energy at phase transition points takes
place. Thus, it is possible to define a new criterion of
transition to chaos. The electrostatic potential can be regarded
as order parameter for the transition.

We conclude that complex generalization of approximate RG method
appears to be useful to advance understanding of the critical
phenomena at threshold of chaos and for development of analogy
with the theory of phase transition, which can give new approaches
to investigation of these phenomena.

\section{Acknowledgements}
The authors acknowledge support from Research Educational Center
of Nonlinear Dynamics and Biophysics at Saratov State University
(REC-006) and RFBR (grant No. 03-02-16074).

\end{document}